\documentclass[aip,reprint,amsmath,amssymb]{revtex4-1}

\usepackage{graphicx}
\usepackage{dcolumn}
\usepackage{bm}
\usepackage{amsmath}

\begin{document}
\title{Low frequency resistance and critical current fluctuations in Al-based Josephson junctions}

\author{C. D. Nugroho}
\email{nugroho2@illinois.edu}
\affiliation{Department of Physics and Materials Research Laboratory, University of Illinois at Urbana-Champaign, Urbana, Illinois 61801, USA}

\author{V. Orlyanchik}
\affiliation{Department of Physics and Materials Research Laboratory, University of Illinois at Urbana-Champaign, Urbana, Illinois 61801, USA}

\author{D. J. Van Harlingen}
\affiliation{Department of Physics and Materials Research Laboratory, University of Illinois at Urbana-Champaign, Urbana, Illinois 61801, USA}


\begin{abstract}
We present low-temperature measurements of the low-frequency $1/f$ noise arising from an ensemble of two-level fluctuators in the oxide barrier of Al/AlO$_{x}$/Al Josephson junctions. The fractional noise power spectrum of the critical-current and normal-state resistance have similar magnitudes and scale linearly with temperature, implying an equivalence between the two. Compiling our results and published data, we deduce the area and temperature scaling of the noise for AlO$_{x}$ barrier junctions. We find that the density of two-level fluctuators in the junction barrier is similar to the typical value in glassy systems. We discuss the implications and consistency with recent qubit experiments.
\end{abstract}

\maketitle


Recent progress in superconducting qubits have resulted in longer coherence times. How far this improvement can continue depends crucially on the losses intrinsic to the Josephson junction. Qubit energy spectroscopy have revealed a density of avoided level crossings arising from the interaction of the qubit with two-level systems (TLSs) in the junction barrier~\cite{BrittonFluxSplittings, MicahTLS, UstinovTLSCoupling}. Additionally, critical-current fluctuations in Josephson junctions are known to exhibit a low-frequency $1/f^{\alpha}$ ($\alpha \sim 1$) spectrum~\cite{Savo,FogliettiSQUIDs}, which is generally understood to arise from a collection of TLSs in the tunnel barrier~\cite{DuttaHornModel, RogersBuhrmanConductanceFluctuations}. The precise microscopic origin of these TLSs and the coupling mechanisms remain relatively unknown.

A survey of a variety of junction architectures from several laboratories found the low frequency $1/f$ critical-current noise power spectral density, $S_{I_{c}}$, to have an almost universal magnitude at $T=4.2~\textrm{K}$ ~\cite{DVHUniversal}. By postulating a $T^{2}$ temperature dependence based on noise measurements in dc superconducting quantum interference devices (SQUID)~\cite{WellstoodT2} and charge qubits~\cite{ChargeQubitNoiseTDependence}, an almost universal noise characteristic was proposed: $S_{I_{c}}/I_{c}^{2} (f=1~\textrm{Hz}) \approx 1.44\times 10^{-10}(T/4.2~\textrm{K})^{2}~\textrm{Hz}^{-1}$.

However, results on the tunneling resistance noise~\cite{Eroms} $S_{R_{n}}/R_{n}^{2}$ in Al/AlO$_{x}$/Al shadow junctions showed a linear temperature dependence and an equivalent critical-current noise magnitude at $4.2~$K that is three orders of magnitude lower than previous measurements. The comparison between resistance and critical current noise is made through the Ambegaokar-Baratoff relation, $I_{c}R_{n}=\pi\Delta/2 \tanh{\left(\frac{\Delta}{2k_{B}T}\right)}$, where $\Delta, I_{c}, R_{n}$ are the superconducting energy gap, critical current, and the normal state resistance respectively. This apparent discrepancy between $S_{I_{c}}$ and $S_{R_{n}}$ led to the proposal of a Kondo-traps noise mechanism~\cite{IoffeFaoro1, WilhelmKondo} that could account for the excess noise magnitude and $T^{2}$ dependence in the SC state.

Furthermore, recent measurements of the kinetic inductance noise~\cite{SiddiqiResonatorCriticalCurrent} of Al shadow junctions at $T=25~$mK placed an even lower bound to the noise magnitude than previous measurements of $S_{I_{c}}$ and $S_{R_{n}}$.

To clarify the apparent discrepancy, in this Letter we present extensive measurements of both $S_{I_{c}}$ and $S_{R_{n}}$ in Al/AlO$_{x}$/Al junctions, the material system most relevant in current superconducting qubits. Our measurements show an equivalence between $S_{I_{c}}/I_{c}^{2}$ and $S_{R_{n}}/R_{n}^{2}$ as expected from the Ambegaokar-Baratoff relation. We find a linear temperature dependence in all devices and a noise magnitude that is consistent with previous measurements in Al-shadow junctions~\cite{Eroms}. Our results place an upper limit on the proposed additional Kondo-traps noise contribution. Combining our results with those from other laboratories~\cite{StonyBrookIcNoise,Eroms, JunctionRapidThermalAnnealing} and junction architectures we suggest a scaling relation for noise in AlO$_{x}$ based junctions. We discuss the implications for superconducting qubits and the consistency with recent qubit measurements.


We fabricated shunted and unshunted Al/AlO$_{x}$/Al Josephson junctions using either the double-angle shadow evaporation or the cross technique. In double-angle evaporation the junction electrodes and barrier oxidation are completed in one step without breaking vacuum. In the cross technique the electrodes are deposited in two evaporation steps. In the second evaporation step the base electrode is ion-milled to remove surface contaminants, followed by the controlled oxidation of the junction barrier and the top electrode evaporation. Both junction architectures were completed in a chamber with a base pressure of $\approx 3 \times 10^{-10}~$Torr. For the shunted junctions the shunt resistors ($R_{s}$) were fabricated by e-beam evaporating $60~$nm of Pd, with typical low temperature sheet resistances $R_{s}\approx1.3~\Omega/$sq. The shunt resistors are patterned with large, $300 \times 300~\mu\textrm{m}^{2}$, cooling fins to minimize hot-electron effects~\cite{ClarkeHotElectrons}.

\begin{center}
\begin{figure}[h]
    \includegraphics[width=8.5 cm]{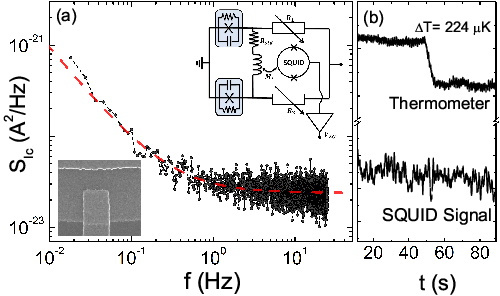}
    \caption{ \label{fig:Fig1}(a) Critical current power spectrum measured at $T=315~\textrm{mK}$, and $I_{b}/I_{c}=1.05$. The dashed line is a fit to the sum of a $1/f$ component which dominates at low frequencies, and a frequency independent component dominating at higher frequencies. Top inset: Schematic representation of the bridge-SQUID potentiometry circuit. Bottom inset: SEM image of the Al/AlO$_{x}$/Al junctions. (b) Simultaneous time traces of the sample temperature and the SQUID signal, monitored across a small temperature perturbation at $T=315~\textrm{mK}$ for $I_{b}/I_{c}=1.05$.}
\end{figure}
\end{center}

Measurements of the critical current noise $S_{I_{c}}$ were performed in a bridge configuration (Fig.~\ref{fig:Fig1}(a) inset) with a dc SQUID monitoring the current fluctuations across the two bridge arms. A small $R_{std}\approx 0.5~\Omega$ is placed in series with the SQUID pickup loop. The two shunted junctions in the bridge are matched and fabricated on-chip in the same lithography step. Potentiometers at room-temperature are used to adjust the currents through the junctions, while the dc-SQUID monitored the voltage imbalance, which is typically kept at zero.

Since the critical-current noise in the junctions are uncorrelated, their noise power contributions add to give the total noise seen at the SQUID:

\begin{multline}
S_{I}^{sq}=\dfrac{R_{D1}^{2}}{(R_{std}+R_{D1}+R_{D2})^{2}}\left(\dfrac{I_{c1}}{I_{1}} \right)^{2}S_{I_{c1}}+ \\ \dfrac{R_{D2}^{2}}{(R_{std}+R_{D1}+R_{D2})^{2}}\left(\dfrac{I_{c2}}{I_{2}} \right)^{2}S_{I_{c2}},
\end{multline}

where $I_{i}$, $I_{ci}$, and $R_{Di} = \partial V_{i}/\partial I_{i}$ are the bias current, critical current and the dynamic resistance of the $i$th junction respectively. When the junctions are closely matched, $R_{D1}\approx R_{D2} = R_{D}$, and are biased near $I_{c}$, $I_{c}/I_{b}\approx 1$, the noise seen by the SQUID is given by:

\begin{equation}
    S_{I}^{sq} = \frac{1}{2}S_{I_{c}},
\end{equation}
where $S_{I_{c}} = \frac{1}{2}(S_{I_{c1}}+S_{I_{c2}})$, is the averaged critical-current noise power density across the two junctions. The reduction by a factor of two in the noise power sensitivity compared to the standard SQUID potentiometry technique~\cite{WellstoodT2} is compensated by the large attenuation of common-mode noise sources such as spurious temperature fluctuations and external biasing noise.

In double-angle evaporated junctions the critical currents can be matched to within $1\%$, allowing for a high attenuation of small temperature fluctuations $\sim 100$s $\mu$K (Fig.~\ref{fig:Fig1}(b)). The residual signal from common temperature fluctuations can be detected when the fluctuations exceed $\sim 1~$mK, which is much larger than the typical thermal instabilities of the system (examined in detail in Ref.~\onlinecite{AntonTemperatureFluctuations}).

The system background noise was determined by monitoring the SQUID output while keeping the junctions in the superconducting state (zero bias). In this regime the high-frequency, $f>10~\textrm{Hz}$, noise power spectrum is dominated by the Johnson noise of the standard resistor, while the low-frequency, $f<10~\textrm{Hz}$ power spectrum is dominated by the $1/f$ flux noise of the SQUID and feedback electronics. A base $1/f$ equivalent flux noise of $S_{\Phi}^{1/2}(1~\textrm{Hz}) \approx 6~\mu\Phi_{o}/\sqrt{\textrm{Hz}}$ is observed, consistent with the calibration data for the SQUID sensor. The background $1/f$ noise is subtracted from the measured data and the remainder is attributed to fluctuations in the junctions.

\begin{center}
\begin{figure}[h]
    \includegraphics[width=8.5cm]{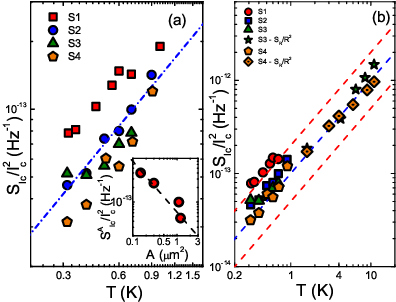}
    \caption{\label{fig:Fig2}(a) Temperature dependence of the critical-current noise fractional power spectral density at $1~$Hz, normalized to a junction area, $A=1~\mu\textrm{m}^{2}$. The dashed line shows the averaged magnitude and a linear temperature dependence. Inset: The area dependence of the critical-current noise fractional power spectral density at $500~$mK and $1~$Hz (not area normalized). The dashed line shows a $1/A$ dependence. (b) Temperature dependence of the critical-current and resistance noise fractional power spectral density at $1~\textrm{Hz}$ (area normalized). The dashed lines are the average, upper, and lower bounds of the resistance noise measured in unshunted junctions.}
\end{figure}
\end{center}

Fig.~\ref{fig:Fig1}(a) shows the critical-current noise power spectral density for device S1 taken at $T=315~$mK and $I_{b}=10~\mu$A, corresponding to $I_{b}/I_{c}\approx 1.05$. The low-frequency power spectrum shows a $1/f^{\alpha}$ noise which crosses over to the frequency independent noise resulting from the thermal noise of the shunt and the mixed down junction quantum noise~\cite{DVHQuantumNoise}. It is important to mention that for all the junctions reported here the measured values of $\alpha$ ranged between 0.9 to 1.1 and were independent of T. The dependence of the noise fractional power spectral density on the junction area is shown in Fig.~\ref{fig:Fig2}(a) (inset). The inverse area scaling ($1/A$) of the fractional power spectral density is consistent with noise resulting from an ensemble of uncorrelated fluctuators, with sufficient density to produce featureless $1/f$ spectrum in all measured junctions down to $\approx 0.1~\mu\textrm{m}^{2}$. Furthermore, the $1/A$ dependence rules out spurious fluctuations induced by the bath temperature, which would have generated a noise that is independent of the junction area~\cite{AntonTemperatureFluctuations}.

The critical-current noise fractional power spectral density ($S_{I_{c}}/I_{c}^{2}$), increases linearly with temperature (Fig.~\ref{fig:Fig2}(a)) with the average T dependence given by $S_{I_{c}}/I_{c}^{2} \approx 1.3\times 10^{-13}~(T/1~\textrm{K})~\textrm{Hz}^{-1}$ (Fig.~\ref{fig:Fig2}(a), dashed line). Both, the noise magnitude and the linear-$T$ dependence are consistent with results in Nb/AlO$_{x}$/Nb trilayer junctions~\cite{StonyBrookIcNoise}, but differed from the $T^{2}$ dependence observed in dc-SQUIDs~\cite{WellstoodT2}.

Following the theoretical proposals in Ref.~\onlinecite{IoffeFaoro1, WilhelmKondo}, the noise characteristics in the superconducting state and normal state are expected to be different. To directly observe this possible crossover we extended the noise measurement for samples S3 and S4 above the $T_{c} \sim 1.25~\textrm{K}$ of Al. For $T>T_{c}$ the shunted junction is represented by the equivalent parallel resistance $R_{eq} = R_{n}R_{s}/(R_{n}+R_{s})$, of $R_{s}$ the shunt resistance, and $R_{n}$ the tunneling resistance the unshunted junction. Assuming that the noise is dominated by fluctuations in $R_{n}$ (fluctuations in $R_{s}$ are small), we can relate the tunneling resistance noise power spectral density $S_{R_{n}}/R_{n}^{2}$ to the signal measured by the SQUID $S_{I}^{sq}$ as:
\begin{equation}
    \frac{S_{R_{n}}}{R_{n}^{2}} = \frac{1}{2}\left(\frac{R_{\Sigma}}{R_{n}}\right)^{2}\frac{1}{(dR_{eq}/dR_{n})^{2}} \frac{S_{I}^{sq}}{I^{2}},
\end{equation} where $R_{\Sigma}$ is the total resistance in the loop, which can be determined from the Johnson-Nyquist noise measured at zero bias, and $I$ is the current bias through each of the junctions. Since $R_{s} < R_{n}$, most of the current flows through the shunt resistor, thus to measure fluctuations due to the tunneling resistance we used bias currents $I \approx 50 - 250~\mu$A to obtain sufficient signal visibility. To remove the ambiguity from self-heating we verified the quadratic dependence of the noise power on biasing current $S_{I}^{sq}\propto I^{2}$.

The measured $S_{R_{n}}/R_{n}^{2}$ varies linearly with temperature similar to the dependence in $S_{I_{c}}/I_{c}^{2}$, and with a noise power magnitude consistent with the equivalence $S_{I_{c}}/I_{c}^{2} = S_{R_{n}}/R_{n}^{2}$ as expected from the Ambegaokar-Baratoff relation. Fig.~\ref{fig:Fig2}(b) shows the comparison between $S_{R_{n}}/R_{n}^{2}$ and $S_{I_{c}}/I_{c}^{2}$. The dashed lines in Fig.~\ref{fig:Fig2}(b) are the average, lower, and upper bounds of the resistance noise measured in unshunted junctions, which reinforces the equality between critical-current and resistance noise.

To rule out a contribution of the low-frequency noise from the shunt resistors we have also measured $S_{R_{n}}/R_{n}^{2}$ on a collection of unshunted junctions. The measurement of the resistance fluctuations were performed in an ac-bridge configuration as described in Ref.~\onlinecite{Eroms}. We used first stage amplifiers with an input noise $\approx 1.6~\textrm{nV}/\sqrt{\textrm{Hz}}$ for a $1~\textrm{k}\Omega$ load impedance. For some of the junctions we have also measured the noise without the bridge layout, while still computing the cross-spectral density of the two readout branches. The modulation frequency is dictated by the sample and setup resistance and capacitance and is typically in the range $1-3~\textrm{kHz}$. Measurements below the critical temperature $T_{c}$ of Al were done by suppressing the superconductivity with an applied magnetic field, $B_{\bot}>100~$mT.

\begin{figure}[h]
    \includegraphics[width=8.5cm]{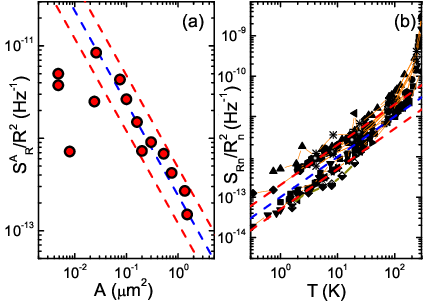}
    \caption{\label{fig:Fig3}(a) Area dependence of the resistance noise fractional power spectral density at $1~$Hz in unshunted junctions at $T=2~\textrm{K}$. Note the deviation from the linear scaling at $A\sim 0.04~\mu\textrm{m}^2$. (b) Temperature dependence of the resistance noise fractional power spectral density at $1~$Hz in unshunted junctions, normalized to $A=1~\mu\textrm{m}^{2}$. The dashed lines depict the average noise magnitude as well as its upper and lower bounds.}
\end{figure}

The area dependence of the resistance noise fractional power spectral density in unshunted junctions (Fig.~\ref{fig:Fig3}(a)) follows $S_{R_{n}}/R_{n}^{2} \propto 1/A$, suggesting the averaging over an ensemble of uncorrelated fluctuators. The area scaling and a featureless $1/f$ power spectrum are found to hold down to $A\approx 0.04~ \mu \textrm{m}^{2}$ at $T=2~\textrm{K}$, indicating a high degree of barrier uniformity. For junction areas in the limit $A \leq 0.04~\mu\textrm{m}^{2}$ at $T=2~\textrm{K}$ the noise becomes more non-gaussian leading to a deviation from the linear area scaling. At even smaller junction areas the noise is dominated by distinct two level systems, where the noise fractional power spectral density at $1~$Hz is generally lower than that expected from the $\propto 1/A$ scaling. The detailed dynamics of the fluctuator and transition to $1/f$ noise is beyond the scope of this letter and will be discussed elsewhere.

Fig.~\ref{fig:Fig3}(b) shows the dependence of the area normalized $S_{R_{n}}/R_{n}^{2}$ on temperature. The three dashed lines in Fig.~\ref{fig:Fig3}(b) trace out the upper and lower bounds, and the average magnitude of the noise. The average noise magnitude over all the junctions is well fitted by $S_{R_{n}}^{\textrm{av}}/R_{n}^{2}=1\times 10^{-13}~(T/1~\textrm{K})~\textrm{Hz}^{-1}$, while the upper and lower bounds differ by a factor of two from this value. Part of the spread can be explained by the uncertainty in the junction sizes. The averaged tunneling-resistance noise and the observed upper and lower limits are consistent with the measured critical-current noise as shown by the dashed lines in Fig.~\ref{fig:Fig2}(b). For $T\approx 70~\textrm{K}$ we observed a deviation from the linear temperature scaling, which may indicate the thermal activation of additional noise mechanisms.


Among the important questions is the location of the fluctuating TLSs. We note that recent noise measurements in junctions with AlO$_{x}$ barriers and Nb electrodes~\cite{StonyBrookIcNoise} showed similar noise magnitude and a linear temperature dependence. Additionally, we measured identical noise characteristics in double-angle evaporated AlO$_{x}$ junctions with $\sim 1~\textrm{nm}$ of Ag deposited on the AlO$_{x}$-Al interface. Lastly, the same noise behavior is observed even in the AlO$_{x}$ cross junction where the barrier is oxidized after an aggressive ion milling of the surface. These properties suggest that the noise sources are insensitive to the barrier interfaces and that the main contribution comes from TLSs buried within the amorphous AlO$_{x}$ barrier. Most likely the TLSs correspond to atoms that tunnel between two positions in the barrier modifying the local tunneling probability. Combining our results with measurements from different laboratories~\cite{Eroms, JunctionRapidThermalAnnealing} and various junction architectures~\cite{StonyBrookIcNoise} lead to the empirical formula for T dependence of the $1/f$ noise in AlO$_{x}$ based junctions:
\begin{equation}
    \frac{S_{R_{n}}}{R_{n}^{2}} = \frac{S_{I_{c}}}{I_{c}^{2}} \approx \frac{1}{A/\mu m^{2}}\left(\frac{T}{1~\textrm{K}}\right)\times \frac{1}{f} \times 10^{-13}~\textrm{Hz}^{-1}
\end{equation}

The low-frequency noise properties measured here can be used to estimate the density of TLSs in the amorphous AlO$_{x}$. Gaussian featureless $1/f$ spectrum requires the averaging of $\sim 1-2~$ active fluctuators per frequency octave~\cite{Restle}. In our measurements we observed a crossover from featureless $1/f$ and the onset of non-gaussianity for junctions with $A\leq 0.04~\mu\textrm{m}^{2}$ at $T=2~\textrm{K}$. Assuming that the $1/f$ noise persists in the frequency range $1~\mu\textrm{Hz}-1~\textrm{THz}$ ($\sim 60~$octaves), the estimated density of fluctuators is $\rho_{TLS}\sim 10^{17}-10^{18}~\textrm{cm}^{-3}~\textrm{K}^{-1}$. This estimate is similar to the almost universal density of TLSs in glassy systems~\cite{ClareYuSaysStudyTLS, WAPhillipsTLSGlasses}, and consistent with the density of TLSs inferred from the number of avoided level crossings in qubit energy spectroscopy~\cite{MicahTLS,MartinisDielectricLoss,PappasEpitaxialQubits} -  $\sim 0.5~\textrm{GHz}^{-1}~\mu\textrm{m}^{-2}$.

\begin{figure}[h]
    \includegraphics[width=8.5cm]{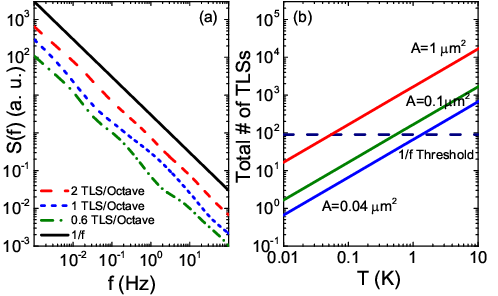}
    \caption{\label{fig:Fig4}(a) Simulated noise power spectrum resulting from averaging over different number of TLSs per frequency octave. (b) Amount of fluctuators extracted from the measured $1/f$ noise in AlO$_{x}$ junctions, as a function of temperature for several junction areas. The dashed line shows the statistical threshold for featureless $1/f$ noise.}
\end{figure}

Fig.~\ref{fig:Fig4}(b) plots the total number of active TLSs in the barrier as a function of temperature assuming the estimated TLS density. The threshold which may be relevant for qubit architectures is set by having only a few TLSs in the entire tunnel barrier. For temperatures $T\approx 50~\textrm{mK}$ and a junction size $A\sim 0.01~\mu\textrm{m}^{2}$, which is the size of the junctions in transmon~\cite{SchreierTransmon} and flux qubits~\cite{BrittonFluxSplittings}, the average number of TLSs is estimated to be less than one. This size threshold is consistent with the high intrinsic quality factor of the junctions observed in 3D transmon~\cite{3DTransmon} and measurements of the junction kinetic inductance noise~\cite{SiddiqiResonatorCriticalCurrent}. Additionally, recent measurements of the free induction decay in 3D Transmons have shown a $\sim 15$~kHz beating in the qubit frequency, which would be consistent with the presence of one active TLS in the junction~\cite{SchoelkopfPrivate}.

Following the arguments of Ref.~\onlinecite{IoffeFaoro1} we derive:
\begin{equation}
    \frac{S_{I_{c}}}{I_{c}^{2}}(f) \approx (\delta A)^{2}\rho t \frac{1}{A} \frac{T}{f},
\end{equation}
where $\delta I_{c}=I_{c}(\delta A/A)$, $\delta A$ parameterizes the effective change in the area of the junction, $t \approx 1-2~\textrm{nm}$ is the tunnel barrier thickness and $\rho$ is the density of TLSs. Using the density of TLSs obtained previously and the measured noise magnitude, we estimate the effective change in the area of the junction to be $\delta A \sim 0.1 - 0.3~\textrm{nm}^{2}$, consistent with Ref.~\onlinecite{IoffeFaoro1, Eroms}.

In summary, we have measured both the critical current and tunnel resistance noise in Al/AlO$_{x}$/Al Josephson junctions, the material most commonly used in superconducting qubits. The measurements uphold the equivalence $S_{I_{c}}/I_{c}^{2} = S_{R_{n}}/R_{n}^{2}$ with a linear temperature dependence down to the lowest temperatures measured. We observed a breakdown of the $1/f$ noise scaling at small junction areas, which gives an estimated TLS density consistent with observations from qubit energy spectroscopy and glassy systems.

This research was funded by the Office of the Director of National Intelligence (ODNI), Intelligence Advanced Research Projects Activity (IARPA), through the Army Research Office. All statements of fact, opinion, or conclusions contained herein are those of the authors and should not be construed as representing the official views or policies of IARPA, the ODNI, or the U.S. Government.

\end{document}